\documentclass[aps,prc,reprint,superscriptaddress,nofootinbib,nolongbibliography]{revtex4-2}

\usepackage[utf8]{inputenc}
\usepackage{CJK}
\usepackage{hyperref,graphicx,amsmath,braket,xcolor,orcidlink}

\begin{document}
\begin{CJK*}{UTF8}{gbsn}

\title{Multiple shape coexistence near $^{118}$Sn: First  $0^+_3$ lifetime measurement}

\author{F.~Wu (吴桐安)~\orcidlink{0000-0003-2739-2377}} \email[]{twa73@sfu.ca}
\affiliation{Department of  Chemistry, Simon Fraser University, Burnaby, BC, V5A 1S6, Canada}
\author{C.R.~Ding~\orcidlink{0009-0001-7243-1518}} 
\affiliation{School of Physics and Astronomy, Sun Yat-sen University, Zhuhai, 519082, China}
\author{C.~Andreoiu~\orcidlink{0000-0002-0196-2792}} 
\affiliation{Department of  Chemistry, Simon Fraser University, Burnaby, BC, V5A 1S6, Canada}
\author{V.~Karayonchev~\orcidlink{0000-0002-6497-0175}}
\altaffiliation[Present address: ]{GSI Helmholtzzentrum f\"{u}r Schwerionenforschung GmbH, Darmstadt, 64291, Germany
}
\affiliation{Argonne National Laboratory, Argonne, IL, 60439, USA}
\author{Y.~Li~\orcidlink{0009-0001-5334-5374}}
\affiliation{School of Physics and Astronomy, Sun Yat-sen University, Zhuhai, 519082, China}
\author{C.~Michelagnoli~\orcidlink{0000-0002-2661-6562}}
\affiliation{Institut Laue-Langevin, Grenoble, F-38042, France}
\author{C.M.~Petrache~\orcidlink{0000-0001-8419-1390}}
\affiliation{Universit\'e Paris-Saclay, CNRS/IN2P3, IJCLab, Orsay, 91405, France}
\affiliation{Institute of Modern Physics, Chinese Academy of Sciences, Lanzhou, 730000, China}
\affiliation{Department of  Chemistry, Simon Fraser University, Burnaby, BC, V5A 1S6, Canada}
\author{J.-M.~R\'egis~\orcidlink{0009-0002-3982-3285}}
\affiliation{Institut f\"ur Kernphysik, Universit\"at zu K\"oln, K\"oln, 50937, Germany}
\author{J.M.~Yao~\orcidlink{0000-0001-9505-1852}}
\affiliation{School of Physics and Astronomy, Sun Yat-sen University, Zhuhai, 519082, China}
\author{M.~Beuschlein~\orcidlink{0009-0001-3183-7503}}
\affiliation{Technische Universit\"at Darmstadt, Department of Physics,
Institute for Nuclear Physics, 64289 Darmstadt, Germany}
\author{G.~Colombi~\orcidlink{0000-0002-8053-9326}}
\altaffiliation[Present address: ]{Department of Physics, University of Guelph, Guelph, ON, N1G 2W1, Canada}
\affiliation{Institut Laue-Langevin, Grenoble, F-38042, France}
\author{J.M.~Daugas}
\author{L.~Domenichetti~\orcidlink{0009-0005-5953-7405}}
\affiliation{Institut Laue-Langevin, Grenoble, F-38042, France}
\author{A.~Esmaylzadeh~\orcidlink{0000-0003-0408-3774}}
\affiliation{Institut f\"ur Kernphysik, Universit\"at zu K\"oln, K\"oln, 50937, Germany}
\author{P.E.~Garrett~\orcidlink{0000-0002-2069-1350}}
\affiliation{Department of Physics, University of Guelph, Guelph, ON, N1G 2W1, Canada}
\affiliation{Department of Physics, University of the Western Cape, P/B X17, Bellville ZA-7535, South Africa}
\author{J.~Jolie~\orcidlink{0000-0003-4617-3786}}
\author{M.~Ley~\orcidlink{0000-0003-1453-070X}}
\affiliation{Institut f\"ur Kernphysik, Universit\"at zu K\"oln, K\"oln, 50937, Germany}
\author{S.~Pannu~\orcidlink{0009-0007-3897-5187
}}
\affiliation{Department of Physics, University of Guelph, Guelph, ON, N1G 2W1, Canada}
\author{P.~Spagnoletti~\orcidlink{0000-0002-7674-989X}}
\altaffiliation[Present address: ]{Department of Physics, University of Liverpool, Liverpool, L69 7ZE, UK}
\affiliation{Department of  Chemistry, Simon Fraser University, Burnaby, BC, V5A 1S6, Canada}
\author{E.~Taddei~\orcidlink{0009-0005-3482-6542}}
\affiliation{Department of  Chemistry, Simon Fraser University, Burnaby, BC, V5A 1S6, Canada}

\date{\today}

\begin{abstract}
The intruder bands in Sn isotopes, built on the 2p-2h excitation across the $Z = 50$ proton shell gap, are well-known examples of shape coexistence near the neutron mid-shell region. Spectroscopic signatures for shape coexistence include enhanced $E0$ transitions between the $0^+$ band heads. However, the underlying shape coexistence and mixing has been unclear because lifetime information for the excited $0^+$ states was incomplete in $^{118}$Sn. We thus present here the first measurement of the $0^+_3$ lifetime in $^{118}$Sn using the fast-timing technique following thermal-neutron capture. The observed enhancement in $\rho^2(E0; 0^+_3 \rightarrow 0^+_2)$ of 150(30) milliunits provides compelling indications for multiple shape coexistence in $^{118}$Sn. Additionally, three distinct shapes in $^{116,118,120}$Sn naturally emerged in theoretical calculations based on the quantum-number-projected generator coordinate method employing a relativistic energy density functional. 
\end{abstract}
\maketitle
\end{CJK*}

\textit{Introduction}\textemdash
The $Z=50$ semimagic Sn isotopes, which spanan across two major neutron shell closures, constitutes one of the best-studied isotopic chains in the nuclear chart. With ten stable isotopes, the most of any element, the Sn isotopes have been studied by a wide variety of probes resulting in a rich set of spectroscopic data that is ideal for nuclear structure investigations.\par 
A surprising yet intriguing feature of the Sn nuclei is 
shape coexistence, where states in the same nucleus exhibit distinct shapes. Excited $J^\pi = 0^+$ states around 2~MeV were observed in the even $^{112-118}$Sn isotopes via two-proton transfer reactions which could not be explained by the spherical vibrator model~\cite{FIELDING1977389}. Since these excited $0^+$ states were populated with significantly higher cross section via the two-proton transfer than the two-neutron transfer~\cite{Fleming1970}, they are interpreted to arise from proton two-particle two-hole~(2p-2h) excitations~\cite{FIELDING1977389}.  
Following the subsequent discovery of collective bands built on these excited $0^+$ states~\cite{BRON1979335}, a second prolate shape was suggested to be present in the neutron mid-shell $^{112-118}$Sn~\cite{BRON1979335}, in addition to the spherical shape traditionally expected due to the $Z=50$ shell closure. These bands were explained as being rotational in nature built on the deformed proton 2p-2h configurations that intrude into the low excitation energy region of spherical states dominated by neutron excitations.\par

Complementary to the transfer reaction~\cite{FIELDING1977389,Fleming1970} and fusion-evaporation~\cite{BRON1979335} studies, the electromagnetic transition rates in the even-even $^{112-124}$Sn isotopes were systematically investigated by the collaboration of B\"acklin~{\it et al.}~\cite{Bcklin1981,Jonsson1981,Kantele1979}. In particular, the observed strong $E0$ transition in $^{116}$Sn with $\rho^2(E0;0^+_3\rightarrow0^+_2) = 100(20)$ milliunits~\cite{Bcklin1981} indicated strong mixing of different shapes with deformation difference $\Delta\beta_2>0.22$ between the intruder $0^+_2$ and $0^+_3$ states~\cite{Kantele1979}. At the time, the large $\Delta\beta_2$ was simply discussed as the coexistence of one spherical and one deformed shape~\cite{Kantele1979}.\par

Very recently, we observed a remarkably large $\rho^2(E0;0^+_3\rightarrow0^+_2)$ value of 120(50)~milliunits also in $^{120}$Sn~\cite{PhysRevC.111.L051307}, which provides direct experimental evidence for shape coexistence and strong mixing between the excited $0^+_2$ and $0^+_3$ states in $^{120}$Sn~\cite{PhysRevC.111.L051307}, similar to the case of $^{116}$Sn~\cite{Kantele1979}. While such $\rho^2(E0)$ transition strengths are direct probes of the shape difference and mixing amplitudes~\cite{KIBEDI2022103930}, the $\rho^2(E0)$ values from the $0^+_3$ state in $^{118}$Sn have not been known until this work, because the $0^+_3$ lifetime only had  limits of 60~ps~$<\tau(0^+_3)<$~290~ps~\cite{Bcklin1981}. To gain a clearer description of shape coexistence in this region, we performed a direct lifetime measurement of excited states in $^{118}$Sn following thermal neutron capture.\par

Multiple shape coexistence, where three or more shapes appear in the same nucleus, is still considered as a rare phenomenon that has only been suggested in a few nuclei. Recently, this picture has emerged~\cite{Paul_SC,Leoni2024} along the other proton shell closures including in the semimagic Ni~\cite{PhysRevLett.118.162502,PhysRevC.89.031301} and Pb~\cite{Andreyev2000,MontesPlaza2025} isotopes, and in the Cd~\cite{PhysRevLett.123.142502}. We suggest, based on our experimentally determined $\rho^2(E0;0^+_3\rightarrow0^+_2)$ value with the support of the modern energy-density-functional-based generator coordinate method~(GCM)~\cite{Yao:2014PRC} calculations, the novel coexistence of {\it three distinct shapes} in the semimagic $^{118}$Sn and the neighbouring $^{116, 120}$Sn.\par
\par

\textit{Experiment}\textemdash
States in $^{118}$Sn below the neutron separation energy were populated following the $(n,\gamma)$ reaction using thermal neutrons from the research reactor at the Institut Laue-Langevin~(ILL) in Grenoble, France. Thermal neutrons were delivered to the FIssion Product Prompt gamma-ray Spectrometer~(FIPPS)~\cite{FIPPS} experimental station via a reflective neutron guide with a maximum flux of $10^8$~s$^{-1}$cm$^{-2}$. A powder target, isotopically enriched to 89.2\% $^{117}$Sn, was mounted at the centre of FIPPS~\cite{FIPPS}. The target was surrounded by 8 Compton-suppressed high-purity germanium~(HPGe) clover-type detectors positioned at 90$^\circ$, and 15 LaBr$_3$ fast-timing detectors at approximately 45$^\circ$ and 135$^\circ$.
The time difference between LaBr$_3$ detector pairs were recorded using time to amplitude converters~(TACs) through a multiplexed start-stop system~\cite{REGIS201672}.\par
Data was taken for approximately $14$ days of beam on target~\cite{ILLData}, and a total of $1.3\times 10^8$ $\gamma$-ray events were sorted into a $\gamma\gamma\gamma$ cube where two $\gamma$ rays were detected in non-adjacent LaBr$_3$ detectors in coincidence with the 1230-keV $^{118}$Sn $2^+_1\rightarrow0^+_1$ transition in a HPGe detector. The $^{118}$Sn experiment in this work was performed using the identical setup as for $^{120}$Sn~\cite{PhysRevC.111.L051307,WU2025123105}, detailed descriptions of the experimental setup and the data sorting procedure can be found in Refs.~\cite{PhysRevC.111.L051307} and~\cite{WU2025123105}.

\textit{Analysis}\textemdash
The lifetime of the $0^+_3$ state in $^{118}$Sn was extracted using the generalized centroid difference method~\cite{REGIS201672,REGIS2025104152}. The time difference between two LaBr$_3$ signals, recorded by the TAC, was correlated to the $\gamma$-ray energies of the transitions that populated~(feeder) and depopulated~(decay) the $0^+_3$ state. A ``delayed'' event was recorded if the TAC was started by the feeder and stopped by the decay, and an ``anti-delayed'' event was recorded if vice versa. The lifetime of the state, $\tau$, can then be extracted using 
\begin{equation}
    \label{eq:GCD}
    \Delta C = {\rm PRD}(E_\text{feeder},E_\text{decay}) + 2\tau,
\end{equation}
where $\Delta C$ is the centroid difference between the \textit{delayed} and \textit{anti-delayed} time distributions, and ${\rm PRD}(E_\text{feeder},E_\text{decay})$, the prompt response difference~\cite{REGIS2025104152}, is the energy-dependent time response of the fast-timing set up. The PRD in this work was calibrated using feeder-decay pairs from states with known lifetimes in a $^{152}$Eu source and the $^{48}$Ti(n,$\gamma$)$^{49}$Ti reaction. PRD values were determined from the measured $\Delta C$ and literature lifetimes using Eq.~\ref{eq:GCD}. See Ref.~\cite{PhysRevC.111.L051307} and the Supplemental Material~\cite{npjn-xpfj_supp} at [https://journals.aps.org/prc/supplemental/10.1103/npjn-xpfj/Supp.pdf] (and Ref.~\cite{KNAFLA2023168279,VasilPhysRevC.99.024326} therein) for details.\par 

In order to extract the lifetime of the $0^+_3$ state, we use the powerful technique of accounting for the total cascade energy (TCE) from the capture state to the $2^+_1$ state. As thermal neutrons have negligible kinetic energy, all $^{118}$Sn nuclei were populated at an excitation energy equal to $S_n = 9326$~keV~\cite{Wang2012}.
\begin{figure*}[ht]
    \includegraphics[width=\linewidth]{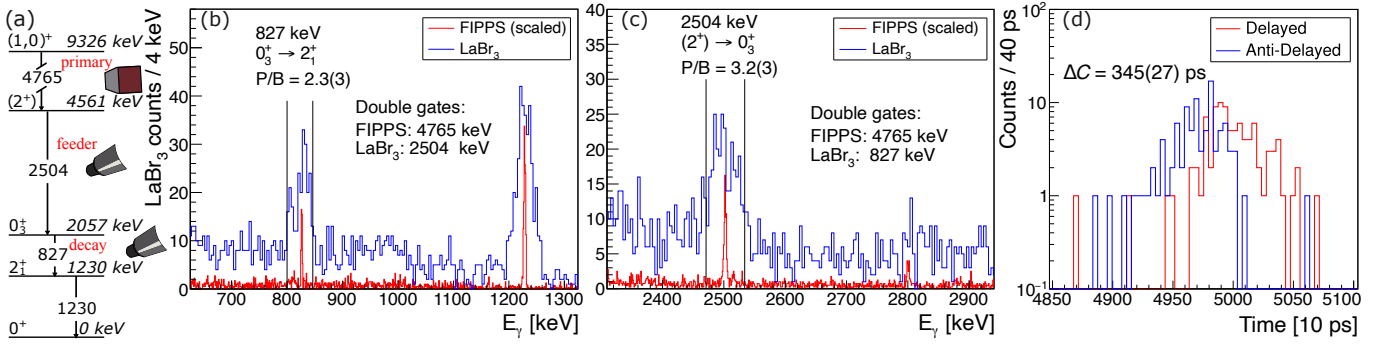}
    \caption{(a) Partial level scheme of $^{118}$Sn with the decay cascade from the capture state at 9326~keV. (b) The LaBr$_3$ spectrum (blue) in coincidence with the 4765-keV primary $\gamma$ ray in HPGe and the 2504-keV feeder in LaBr$_3$ from HPGe-LaBr$_3$-LaBr$_3$ events. Vertical bars represent the LaBr$_3$ gate. The HPGe spectrum (red) under the same gates from HPGe-HPGe-LaBr$_3$ events are shown to resolve features obscured by the LaBr$_3$ width. (c). Same as (b), except with the LaBr$_3$ gate on the 827-keV decay. (d) The delayed and anti-delayed TAC distributions in coincidence with the HPGe-LaBr$_3$-LaBr$_3$ gates shown in (a-c).}
    \label{fig:C4765_L2504}
\end{figure*}
Since the triple-$\gamma$ coincidence accounts for the TCE from the capture state, little background is seen around the 827-keV decay peak in the LaBr$_3$ spectrum with a HPGe gate on the 4765-keV primary $\gamma$ ray and a LaBr$_3$ gate on the 2504-keV feeder, as shown in Fig.~\ref{fig:C4765_L2504}(b). Similarly low background is seen around the 2504-keV feeder peak, as shown in the reverse-gated Fig.~\ref{fig:C4765_L2504}(c).\par

The delayed and anti-delayed TAC distributions in coincidence with a HPGe gate on the primary $\gamma$ ray at 4765~keV, and two LaBr$_3$ gates on the feeder and decay transitions, as illustrated by the vertical bars in Figs.~\ref{fig:C4765_L2504}(b-c), are shown in Fig.~\ref{fig:C4765_L2504}(d). The centroid difference, $\Delta C_\text{exp} = C_\text{delayed} - C_\text{anti-delayed}$ was then determined to be 345(27)~ps. 

Following the Compton-background correction procedure described in Ref.~\cite{REGIS2025104152}, $\Delta C_\text{PP} = 384(31)$~ps was determined as the centroid shift due to the full-energy peaks. Using ${\rm PRD} (2505, 827) =  241(15)$~ps, the lifetime of the $0^+_3$ state was determined with Eq.~(\ref{eq:GCD}), replacing $\Delta C$ with $\Delta C_\text{PP}$,
to be $\tau (0^+_3) = 72(17)$~ps.\par

Two additional cascades with primary $\gamma$ rays of 4407 and 4972~keV were also investigated. These cascades are more complicated because following the primary $\gamma$ ray, the 2057-keV $0^+_3$ state was populated alongside the $\tau(2^+_2) = 4.2(6)$~ps 2043-keV $2^+_2$ state~\cite{Kitao1995} with comparable intensities. Since the $0^+_3$ state is only 14~keV higher in energy than the $2^+_2$, the feeder and decay transitions from these parallel cascades are indistinguishable because of the resolutions of the LaBr$_3$ detectors. Using a weighted-sum procedure to account for the contribution from parallel cascades, which will be discussed in a subsequent publication, the $0^+_3$ lifetime was determined to be 70(25) and 96(36)~ps from the measurements involving the 4407 and 4972-keV cascades, respectively.\par

\begin{figure*}[ht]
    \includegraphics[width=\linewidth]{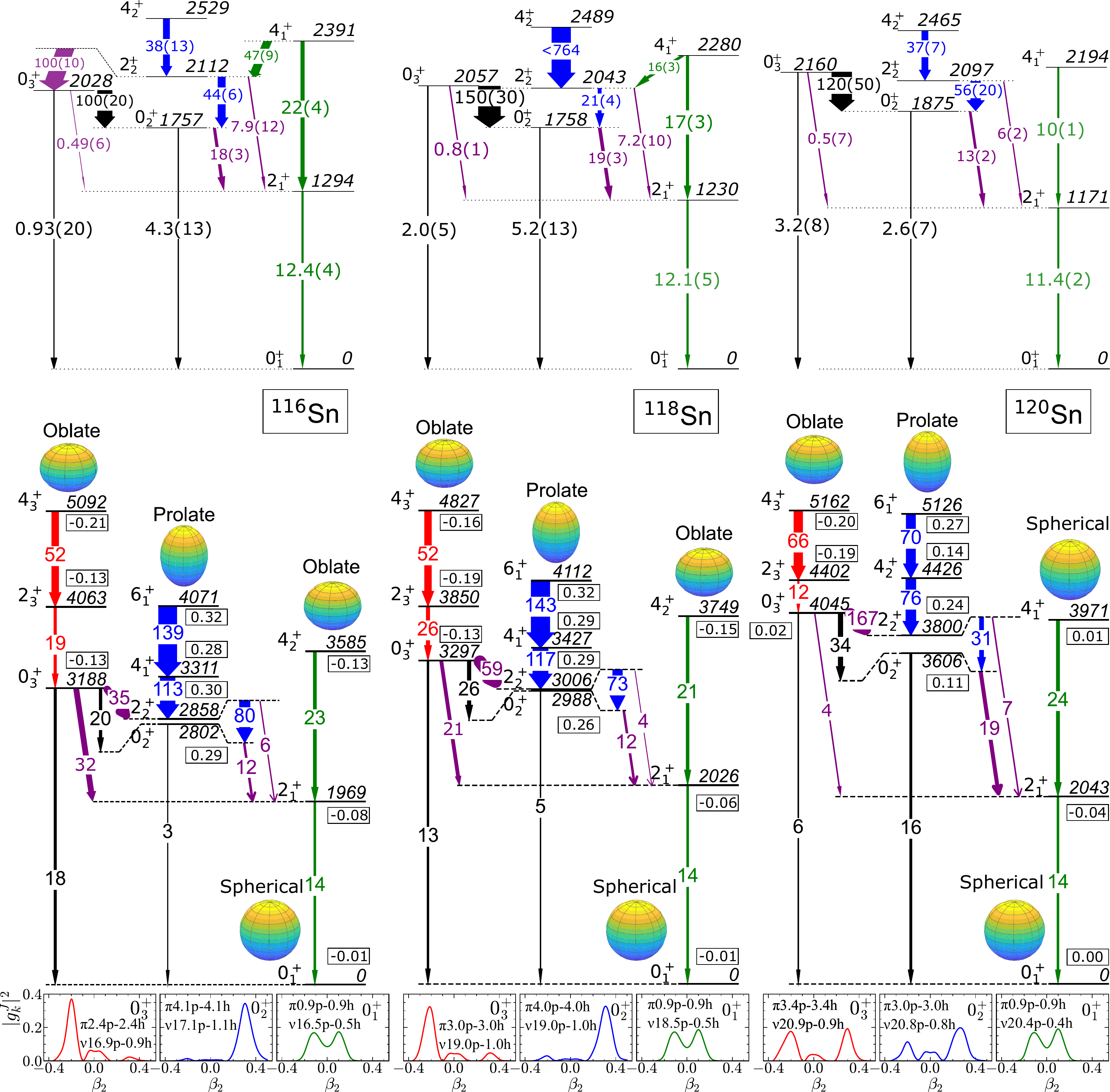}
    \caption{The experimental partial level schemes (upper) for $^{116,118,120}$Sn with the corresponding MR-CDFT calculations (lower). The $B(E2)$ strengths in W.u. are colored in green, blue, and red corresponding to the calculated bands built on the $0^+_1$, $0^+_2$, and $0^+_3$ states, respectively. Purple arrows are interband transitions. Black arrows correspond to $1000\times\rho^2(E0)$ values.  Experimental values from the $0^+_3$ state in $^{118}$Sn are determined using the lifetime from this work, with the rest taken from Refs.~\cite{PhysRevC.99.024303,PhysRevC.102.024323,Bcklin1981,PhysRevC.111.L051307,NNDC,Jonsson1981}.  The collective wave functions $|g^{J}_k(\beta_2)|^2$ are shown below each $0^+$ state. The average quadrupole deformation parameter $\bar{\beta}_2$, defined as $\bar{\beta}_2=\sum_{\beta_2}|g^{J}_k(\beta_2)|^2 \beta_2$, for each state is given inside the box next to the state. The numbers of particles and holes for neutrons ($\nu$) and protons ($\pi$), defined with respect to $^{100}$Sn, for each $0^+$ state are also indicated. See main text for details.
    }
    \label{fig:transition_rates}
\end{figure*}

The $0^+_3$ lifetime is determined by taking the weighted average from the three cascades to be 74(13)~ps. Using the measured lifetime, the $B(E2;0^+_3\rightarrow2^+_1)$ value was calculated to be 0.80(14)~W.u.. Combined with the $X(E0/E2)$ values reported in Refs.~\cite{Bcklin1981} and~\cite{PhysRevC.109.054317}, the $10^3\times \rho^2(E0)$ values for $0^+_3\rightarrow0^+_1$ and $0^+_3\rightarrow0^+_2$ were calculated to be 2.0(5) and 150(30), respectively. The systematics of the experimental electric multipole transition strengths in $^{116, 118, 120}$Sn, including the values obtained in this work, are shown in the top panel of Fig~\ref{fig:transition_rates}.

\textit{Model description}\textemdash
The generator coordinate method (GCM) provides a versatile framework for describing quantum many-body systems, where the total wave function is expressed as a superposition of nonorthogonal basis states generated by continuously varying parameters, the generator coordinates~\cite{Hill:1953,Ring:1980}. It offers an effective means to model large-amplitude collective motion in nuclei. The multireference density functional theory (MR-DFT), which combines the energy density functional approach with quantum-number-projected GCM, has proven highly successful in describing low-lying states, particularly in nuclei exhibiting shape mixing and coexistence~\cite{Bender:2003RMP,Niksic:2011PPNP,Yao:2013PRC,Sheikh:2019}.

In this work, we employ the MR-CDFT~\cite{Yao:2009PRC,Yao:2010,Yao:2014PRC} to study the low-lying states of $^{116,118,120}$Sn. The collective wave functions are constructed as superpositions of symmetry-conserving mean-field states with different quadrupole shapes~\cite{Meng:2005PPNP,Meng:2016Book,Vretenar:2005PR},  and the mixing amplitudes are obtained variationally from the Hill-Wheeler-Griffin (HWG) equation~\cite{Ring:1980}. Solving the HWG equation yields both the excitation energies and collective wave functions of the low-lying states. The definition for the collective wave function  $g^{J=0}_k(\beta_2)$ can be found, for instance, in Refs.~\cite{Yao:2013PRC,Yao:2022_Book}.

 The MR-CDFT results for the energy spectra and electric multipole transition strengths, together with the squares of collective wave functions, are shown in the bottom panel of Fig.~\ref{fig:transition_rates}.  Several low-lying $0^+$ states are obtained which are connected through electric monopole ($E0$) transitions. The corresponding nuclear matrix element is given by
\begin{equation}
\rho^2(E0; 0^+_k \rightarrow 0^+_{k'}) =
\left|
\frac{\braket{0^+_{k'}| e \sum_p r_p^2 | 0^+_k}}{e R^2}
\right|^2,
\label{eq:E0_transition}
\end{equation}
where $R = 1.2A^{1/3}$ fm, and $p$ denotes a proton single-particle state.
Since the matrix elements are evaluated in the full single-particle model space, no effective charge is required, and the bare proton charge is used throughout. Moreover, to compare with the experimental level structures, we have constructed rotational bands based on each $0^+$ state and calculated the intra-band and inter-band $E2$ transition strengths.

\textit{Discussion}\textemdash
 The $E0$ transition strength directly correlates to the size of the deformation and to the amount of mixing between configurations corresponding to different shapes. In general, a large $E0$ strength arises from strong mixing between states with different mean-squared radii~\cite{Heyde:1988PRC,Wood:1999,KIBEDI2022103930}. Indeed, the observed large $10^3\times \rho^2(E0;0^+_3\rightarrow0^+_2) = 150(30)$ in $^{118}$Sn indicates a strong mixture with large shape difference between the $0^+_2$ and $0^+_3$ states, similar to the neighboring $^{116,120}$Sn. Using the method described in Refs.~\cite{Kantele1979} and~\cite{PhysRevC.111.L051307}, the minimum quadrupole shape difference between the two states in $^{118}$Sn can be estimated from just the experimental result to be $\Delta\beta_2 > 0.24$, assuming complete admixture.\par

In the context of the large $\rho^2(E0; 0^+_3\rightarrow0^+_2)$ strengths, we propose in this Letter the coexistence of three distinct shapes in $^{116,118,120}$Sn, as supported by the following experimental observations: i) the ground states of the semimagic Sn nuclei are nearly spherical based on quadrupole moment measurements~\cite{PhysRevC.92.041303}, 
 ii) the intruder 2p-2h bands are deformed~\cite{PhysRevC.99.024303,BRON1979335,FIELDING1977389,Fleming1970,Pore2017}, and iii) there is a large shape difference, as indicated by the large $\rho^2(E0)$ values, between the intruder $0^+_2$ and $0^+_3$ states~\cite{Kantele1979,PhysRevC.111.L051307}.\par 

The coexistence of the three distinct shapes of spherical, prolate, and oblate is also supported by the MR-CDFT calculations, which employ no free parameters adjusted to the experimental excitation energies. Our results are, to some extent, consistent with the potential energy surfaces obtained from MCSM calculations~\cite{Leoni2024,PhysRevLett.121.062501,Corbari2025}, which exhibit both prolate and oblate minima.  To investigate the structure of the first three $0^+$ states in the language of shell models, we computed the occupation numbers of spherical single-particle orbitals by diagonalizing the density matrix of the GCM wave functions~\cite{Zhou:2024vlt,Rodriguez:2016qls}.
The occupation numbers of individual orbitals are provided in detail in the Supplemental Material~\cite{npjn-xpfj_supp} at [URL will be inserted by publisher]. We find that the nearly spherical $0^+_1$ state in all three Sn isotopes can be classified as a $\pi$1p$-$1h excitation across the $Z = 50$ shell. In contrast, the $0^+_2$ states in $^{116,118}$Sn, dominated by prolate shapes, correspond to $\pi$4p$-$4h excitations, and that in $^{120}$Sn corresponds to a $\pi$3p$-$3h excitation. Moreover, from $^{116}$Sn to $^{120}$Sn, the oblate $0^+_3$ state evolves from about $\pi$2p$-$2h to about $\pi$3p$-$3h excitations. These results confirm the conclusions obtained in the previous studies ~\cite{Paul_SC,BRON1979335,PhysRevC.99.024303,PhysRevC.111.L051307,Pore2017,Leoni2024,FIELDING1977389} that the emergence of low-energy intruder states and shape coexistence originates from multi-proton excitations across the $Z = 50$ closed shell.

We note that the MR-CDFT reproduces most of the transition strengths reasonably well, even though the excitation energies are overall overestimated. This stands in marked contrast to shell-model studies employing a $Z = 50$ core~\cite{Holt1998,Ray2022}, which fail to reproduce these strengths. This highlights the essential role of including proton excitations across the $Z = 50$ closed shell and the effect of shape fluctuation in the model calculations.

It is worth pointing out that the overestimation of excitation energies in the MR-CDFT has been usually attributed to the missing of time-odd components of the moving mean field which are expected to significantly compress the entire energy spectrum~\cite{Hinohara:2009}. This issue is expected to be remedied by including either cranking configurations~\cite{Borrajo:2015} or  mixing explicitly noncollective quasi-particle excitation configurations. With this extension, we anticipate a more accurate reproduction of the structure of $0^+_2$ and $0^+_3$ states, leading to the excitation energies and the $\rho^2(E0)$ values that are closer to the experimental data. 

\textit{Conclusion}\textemdash 
In summary, based on the measured $E0$ transition strengths and the MR-CDFT calculations, we suggest multiple shape coexistence in the $Z = 50$ semimagic  $^{116,118,120}$Sn isotopes. Considering observations along other proton-shell closures, such as in Ni and Pb, the phenomenon of multiple-shape coexistence may be more common than currently recognized. 
The suggestion of multiple shape coexistence in the Sn isotopes parallels that of the Cd isotopes.  In the latter, a series of Coulomb excitation studies are actively being undertaken with the goal of extracting a sufficient set of $E2$ matrix elements to form the quadrupole invariant quantities from which the shapes can be inferred on the $(\beta_2,\gamma)$ plane. Such measurements are also needed for the Sn isotopes that would firmly establish the multiple shape coexistence scenario. Furthermore, experiments such as fusion-evaporation with high reaction-channel selectivity are required in order to search for the rotational bands built on the $0^+_3$ states.

\section{Acknowledgments}
The authors thank C.~M\"uller-Gatermann for the target material. F.W. thanks Marco Siciliano for insightful discussions on multiple shape coexistence in Sn and on Coulomb excitation. F.W. acknowledges funding from the Canadian Institute of Nuclear Physics~(CINP) through its Graduate Fellowship program. J.-M.R. and M.L. are funded by the DFG grant JO391 18/2. This work was supported in part by the Natural Sciences and Engineering Research Council of Canada. This material is in part based upon work supported by the U.S. Department of Energy, Office of Science, Office of Nuclear Physics, under Contracts No. DE-AC02-06CH11357 (ANL).
Y.L., C.R.D. and J.M.Y. acknowledge
support from the National Natural Science Foundation of China (Grant Nos. 125B2108, 12405143, 12375119), and the Guangdong Basic and Applied Basic Research Foundation (2023A1515010936).

\section{Author Contribution Statement}
F. W. was the main contributor to the introduction, experiment, analysis, and the experimental discussions, while C. R. D. was the primary contributor to the model calculations and discussion. V.K. led the experimental proposal and planned the experiment with support from C.A., M.B., P.E.G., C.M., C.M.P. and J.-M.R.. C.R.D. and Y.L. performed the MR-CDFT calculations with support from         J.M.Y..  M.B., G.C., J.M.D., L.D., A.E., P.E.G., C.M., V.K, M.L., S.P., J.-M.R., P.S., E.T., and F.W. performed the experiment. C.A., C.M., J.-M.R., J.J., P.E.G., and J.M.Y. acquired financial support for the project. F.W. sorted and analyzed the data with substantial support from V.K, A.E., J.-M.R., and P.S.. F.W. wrote the original manuscript with substantial support from C.R.D., C.A., V.K., C.M.P., and J.M.Y.. J.M.Y. provided supervision to C.R.D. and Y.L., and C.A. supervised F.W.. All authors provided critical feedback and helped shape the research, analysis, and manuscript.


\bibliography{120Sn,MRCDFT}


\end{document}